\documentclass{emulateapj}
\usepackage{apjfonts,epstopdf}

\journalinfo{Astrophysical Journal, in press}
\slugcomment{}

\shorttitle{$L^*$ Progenitors Experiencing Rapid Star Formation}
\shortauthors{Gawiser et al.}

\begin{document}

\title{Ly$\alpha$-Emitting Galaxies at $z=3.1$:  L$^*$ Progenitors 
Experiencing Rapid Star Formation\footnotemark[14]}

\author{Eric Gawiser\footnotemark[1,2,3,4],
Harold Francke\footnotemark[1,3],
Kamson Lai\footnotemark[5],
Kevin Schawinski\footnotemark[6],
Caryl Gronwall\footnotemark[7], 
Robin Ciardullo\footnotemark[7],
Ryan Quadri\footnotemark[1],
Alvaro Orsi\footnotemark[8], 
L. Felipe Barrientos\footnotemark[8],%
Guillermo A. Blanc\footnotemark[3,9],%
Giovanni Fazio\footnotemark[5],%
John J. Feldmeier\footnotemark[10],
Jia-Sheng Huang\footnotemark[5],%
Leopoldo Infante\footnotemark[8],
Paulina Lira\footnotemark[3], 
Nelson Padilla\footnotemark[8],%
Edward N. Taylor\footnotemark[11],
Ezequiel Treister\footnotemark[1,3,12],
C. Megan Urry\footnotemark[1,2,13], 
Pieter G. van Dokkum\footnotemark[1,2],
Shanil N. Virani\footnotemark[1,2]%
}

\email{gawiser@physics.rutgers.edu}

\footnotetext[1]{Department of Astronomy, Yale University, New Haven, CT.}
\footnotetext[2]{Yale Center for Astronomy \& Astrophysics, Yale University, 
New Haven, CT.}
\footnotetext[3]{Departamento de Astronom\'{\i}a, Universidad de Chile, Casilla 36-D, Santiago, Chile.}
\footnotetext[4]{Department of Physics and Astronomy, Rutgers University, 
Piscataway, NJ.}
\footnotetext[5]{Harvard-Smithsonian Center for Astrophysics, Cambridge, MA.}
\footnotetext[6]{University of Oxford, Astrophysics, Keble Road, Oxford UK.}
\footnotetext[7]{Department of Astronomy and Astrophysics, Pennsylvania 
State University, University Park, PA.}
\footnotetext[8]{Departamento de Astronom\'{\i}a y Astrof\'{\i}sica, Pontificia Universidad Cat\'olica de Chile, 
Santiago, Chile.}
\footnotetext[9]{Department of Astronomy, University of Texas at Austin, Austin, TX.}
\footnotetext[10]{Department of Physics \& Astronomy, 
Youngstown State University, Youngstown, OH.}
\footnotetext[11]{Leiden Observatory, Leiden, Netherlands.}
\footnotetext[12]{European Southern Observatory, Santiago, Chile.}
\footnotetext[13]{Department of Physics, Yale University, New Haven, CT.}
\footnotetext[14]{This work is based on archival data obtained
with the Spitzer Space Telescope, which is operated by the Jet Propulsion 
Laboratory, California Institute of Technology under a contract with NASA,  
with the  6.5 m Magellan-Baade 
telescope, 
a collaboration between the Observatories of the Carnegie Institution of 
Washington, University of Arizona, Harvard University, University of 
Michigan, and Massachusetts Institute of Technology, 
and at Cerro Tololo Inter-American 
Observatory, a division of the National Optical Astronomy Observatories, 
which is operated by the Association of Universities for Research in 
Astronomy, Inc. under cooperative agreement with the National Science 
Foundation.}  

\begin{abstract} 

We studied the clustering properties and multiwavelength spectral 
energy distributions of 
 a complete sample of 162 Ly$\alpha$-emitting 
(LAE) galaxies at $z\simeq 3.1$ 
discovered in deep narrow-band 
MUSYC imaging of the Extended Chandra Deep Field South. 
LAEs were selected to have observed frame equivalent widths 
$>$80\AA\ and emission line fluxes 
$>1.5\times10^{-17}$ergs cm$^{-2}$ s$^{-1}$.  
Only 1\% of our LAE sample appears to host AGN.  
The LAEs exhibit a moderate spatial 
correlation length of $r_0=3.6^{+0.8}_{-1.0}$Mpc, 
corresponding
to a bias factor $b=1.7^{+0.3}_{-0.4}$, which implies  
median dark matter halo masses of 
$\log_{10}\mathrm{M_{med}} = 10.9^{+0.5}_{-0.9}$M$_\odot$.
Comparing the number density of LAEs, 
$1.5\pm0.3\times10^{-3}$Mpc$^{-3}$, 
with the number density of 
these halos finds a mean halo occupation $\sim$1--10\%.  
The evolution of galaxy bias with redshift 
implies that most $z=3.1$ LAEs evolve into present-day galaxies
with $L<2.5L^*$, whereas other $z>3$ galaxy populations typically 
evolve into more massive galaxies.    
Halo merger trees show that $z=0$ descendants occupy 
halos with a wide range of masses, with a median descendant 
mass close to that of $L^*$.  
Only 30\% of LAEs have sufficient stellar mass ($>\sim3\times10^9$M$_\odot$) 
to yield detections 
in deep Spitzer-IRAC imaging.  
A two-population SED fit to the stacked
$UBVRIzJK$+[3.6,4.5,5.6,8.0]$\mu$m  
fluxes of the IRAC-undetected 
objects finds that the typical LAE has low stellar mass 
($1.0^{+0.6}_{-0.4}\times10^9$M$_\odot$), 
moderate star formation rate 
($2\pm1 $M$_\odot$yr$^{-1}$), a young component age 
of $20^{+30}_{-10}$Myr, 
 and little dust ($A_V<0.2$).   
The best fit model has 20\% of the mass in the young stellar component, 
but models without evolved stars are also allowed.

\end{abstract}


\section{INTRODUCTION}
\label{sec:intro}

The discovery of high-redshift Ly$\alpha$-emitting galaxies (LAEs)
opened a new frontier in astronomy \citep{cowieh98,huetal98}.
Because the Ly$\alpha$ line is easily quenched, a galaxy with
detectable Ly$\alpha$ emission is likely dust-free, i.e., in
the initial phases of a burst of star formation.  
The Ly$\alpha$ lines have large equivalent widths
($20${\AA}$<$EW$_{\mathrm{rest}}<\sim 100${\AA}) and broad velocity widths 
(100 km~s$^{-1}$$<$FWHM$<$800 km~s$^{-1}$) 
and are often asymmetric, indicative of
high-redshift galaxies undergoing active star-formation 
\citep[e.g.,][]{manningetal00, kudritzkietal00,arnaboldietal02,rhoadsetal03,
dawsonetal04,huetal04,venemansetal05,matsudaetal06,gronwalletal07}.

Other 
high-redshift galaxy 
populations (including Lyman break galaxies, Distant Red Galaxies, Sub-Millimeter Galaxies) exhibit  
strong clustering and should evolve into elliptical and giant elliptical 
galaxies today \citep{adelbergeretal05a,quadrietal07a,webbetal03}.  
These objects 
were selected by unusually strong rest-frame continuum emission in 
the ultraviolet, optical, and far-IR respectively, resulting 
in $10^{10}L_\odot \leq L_{bol} \leq 10^{12}L_\odot$ \citep{reddyetal06b}.
Such strong continua
appear to occur primarily in deep potential wells that are strongly biased 
versus the general distribution of dark matter
halos.  
  LAEs instead offer the chance 
to probe the faint end of the (bolometric) luminosity function of 
high-redshift galaxies, which contains the majority of galaxies.  
The strong Ly$\alpha$ 
emission line allows detection and spectroscopic confirmation of 
LAEs with typical bolometric luminosities 
$\simeq 10^{10} L_\odot$.   
A detailed calculation of the LAE luminosity function at $z\simeq 3.1$
is given in \citet{gronwalletal07}.

Spectral energy distribution (SED) modelling of the stacked $UBVRIzJK$ photometry of 18 LAEs 
in the Extended Chandra Deep Field-South (ECDF-S) 
\citep{gawiseretal06b}
showed the average galaxy to have 
low stellar mass ($<10^9$ M$_\odot$) and minimal dust
\citep[see also][]{nilssonetal07}.
LAEs have the highest {\it specific} star formation rates 
(defined as SFR divided by stellar mass)
of any type of
galaxy, implying the youngest ages \citep{castroceronetal06}.
Because Ly$\alpha$ emission is easily quenched by dust, 
LAEs have often been characterized as protogalaxies experiencing their
 first burst of star formation \citep[e.g.][]{hum96}.  
However, the differing behavior of Ly$\alpha$ and 
continuum photons encountering dust and neutral gas makes it possible for
 older galaxies to exhibit strong 
Ly$\alpha$ emission when morphology and kinematics favor  
the escape of these photons towards Earth
\citep{neufeld91,haimans99,hanseno06}.  
This could allow older, dusty galaxies with actively 
star-forming regions to exhibit Ly$\alpha$ emission with high equivalent
width.  

SED modelling of LAEs using Spitzer-IRAC \citep{fazioetal04}  
to probe rest-frame near-infrared 
wavelengths, where old stars dominate the emission, 
has yielded mixed results.  
\citet{pirzkaletal07} report extremely young ages of a few Myr and low 
stellar masses ($10^6$M$_\odot<$M$_*<10^8$M$_\odot$) from SED modelling of 
9 LAEs at $4.0<z<5.7$ in the Hubble Ultra Deep Field.  
However, \citet{laietal07a}
performed SED fitting to 3 LAEs with
IRAC detections
out of a sample of 12 $z=5.7$ LAEs in GOODS-N and found
ages as high as 700 Myr and 
significant stellar masses ($10^9<$ M$_\odot$ $<10^{10}$), 
making it appear that these LAEs 
were not undergoing their first burst of star formation.
The 9/12 LAEs lacking IRAC detections are presumably 
younger and less massive and might lack an evolved population.  
Investigating the nature of the LAEs without IRAC detections
requires 
a stacking approach to see if
the typical LAE stellar mass is low enough to have been generated
in a single ongoing starburst.  Stacking will yield the best results
when applied to a large statistical sample of LAEs in a region with
deep IRAC imaging.

\citet{gronwalletal07} present 
the largest available sample of LAEs in an unbiased field, 162 LAEs  
at $z=3.1$ in the ECDF-S discovered as part of the MUSYC survey
(\citealp{gawiseretal06a}, \url{http://www.astro.yale.edu/MUSYC}). 
The ECDF-S has been targeted with deep narrow-band imaging 
and multi-object spectroscopy, 
complemented by public broad-band $UBVRIzJK$, Spitzer+IRAC and 
Chandra+ACIS-I imaging.  
We improve the constraints 
of \citet{gawiseretal06b} 
using the MUSYC $UBVRIzJK$ photometry of 
our larger sample of LAEs 
and adding IRAC [3.6$\mu$m,4.5$\mu$m,5.8$\mu$m,8.0$\mu$m] 
Cycle 2 legacy images from 
SIMPLE (Spitzer IRAC/MUSYC Public Legacy of the
ECDF-S, \url{http://www.astro.yale.edu/dokkum/simple}). 

This Letter summarizes our imaging and spectroscopic observations of 
LAEs, presents our results from clustering analysis and SED modelling, 
and discusses the implications for the formation 
process of typical present-day galaxies.  
We assume a $\Lambda$CDM cosmology 
consistent with WMAP results \citep{spergeletal07} with 
$\Omega_m=0.3, \Omega_\Lambda=0.7$,  
$H_0 = 70$ km s$^{-1}$ Mpc$^{-1}$, and rms dark matter fluctuations 
on $8h^{-1}$ Mpc scales given by $\sigma_8 = 0.9$ .  All 
correlation lengths and number densities are comoving.  We have 
suppressed 
factors of $h_{70}$ in reporting correlation lengths, 
number densities, dark matter masses, stellar masses and star 
formation rates.  

\section{OBSERVATIONS}
\label{sec:obs}

\begin{figure}[h!]
\epsscale{1.2}
\plotone{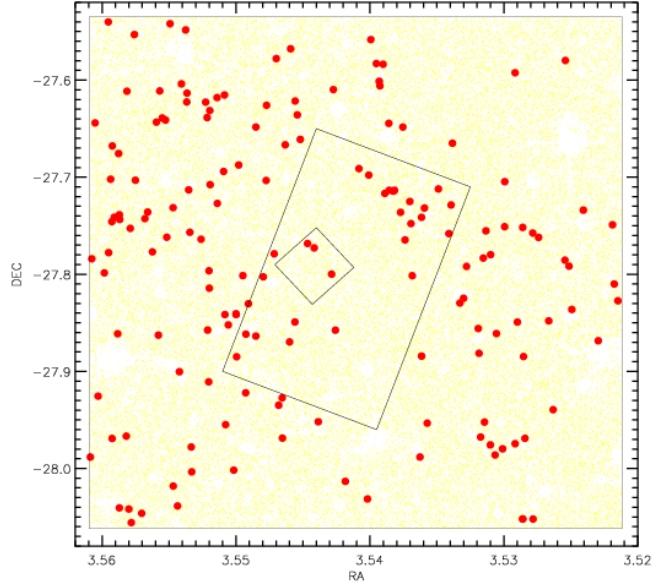}
\caption[]
{ 
Plot of the $31.5' \times 31.5'$ 
Extended Chandra Deep Field-South, showing the 
MUSYC BVR-selected catalog of 84490 objects as tiny dots.  
The rectangle shows the GOODS-S region, 
and the inner square shows the 
Hubble Ultra Deep Field.  
Ly$\alpha$-emitting galaxies at $z=3.1$ are shown as solid circles.
}
\label{fig:ecdfs}
\end{figure}

Our narrow-band 4990{\AA}
and 
$UBVRIzJK$ images of ECDF-S are described in 
\citet{gronwalletal07} and \citet{gawiseretal06b} and are 
available at 
\url{http://www.astro.yale.edu/MUSYC}.
The final images 
cover $31.5'\times31.5'=992$ arcmin$^2$ 
to a narrow-band completeness limit of  $\sim 1.5 \times
10^{-17}$~ergs~cm$^{-2}$~s$^{-1}$ (AB=25.4 in the 50{\AA} 
FWHM NB4990{\AA} filter).
Figure \ref{fig:ecdfs} shows 
our complete sample of 162 strong Ly$\alpha$ emitting galaxies at 
$z\simeq 3.1$ with equivalent width $>$80\AA\ in the observed frame.   
28 LAEs lie in the region surveyed by the GOODS Legacy program
\citep{dickinsonetal03b}.

Multi-object spectroscopy of 92 LAE candidates, along with other MUSYC 
targets,  
was performed with Magellan-Baade+IMACS 
on Oct. 26-27, 2003, Oct. 7-8, 2004, 
Feb. 4-7, 2005, Nov. 2-3, 2005, Oct. 25-27, 2006, Nov. 21-22, 2006 
and Feb. 18-20, 2007.  
The 300 line/mm grism was used with 1.2$''$ slitlets 
to cover $4000-9000$\AA\ at a resolution of 
$R=640$ i.e., 470 km s$^{-1}$, at the wavelength of Ly$\alpha$ emission.  
Mask exposure times ranged from 2 to 5 hours, with the longer exposures 
sufficient to detect Ly$\alpha$ emission lines down to our completeness 
limit of $\sim 1.5 \times10^{-17}$~ergs~cm$^{-2}$~s$^{-1}$, assuming 
clear conditions and minimal slit losses.  
Details of our spectroscopy will be given in P.~Lira et al. (in prep).  
Redshifts were confirmed to lie at $3.08<z<3.14$ for 61 of the LAEs, 
with 1 interloping 
AGN at $z=1.60$ where [C~III]$\lambda$1909 falls 
in the narrowband filter, and
the other 30 objects lacking sufficient S/N to yield redshifts.
Our success rate for 
the slitmasks with the highest S/N was 90\%, setting an upper limit of 10\% on 
possible contamination of our LAE sample.  
The rate of non-detections was higher in masks with shorter exposure times 
resulting from weather or instrument challenges, consistent with the 
reduced S/N.  
 Our spectroscopy shows that the
sample is not contaminated by $z=0.34$ [O II] emission-line galaxies,
which are the typical interlopers for narrow-band-selected LAE
samples.  These have been eliminated by requiring observer's-frame
EW$>80${\AA}
which eliminates all but the 
rarest [O II] emitters (\citealp{terlevichetal91},\citealp{hoggetal98b},\citealp{sternetal00}).   

The Ly$\alpha$ emission in LAEs appears to derive from star formation
rather than AGN activity; only 2/162 objects in our complete sample
are detected as X-ray sources in the Chandra catalogs of CDF-S and
ECDF-S 
\citep{alexanderetal03,lehmeretal05b,viranietal06}.
One is the $z=1.6$ interloper but the other is at $z=3.092$.  
One additional object at $z=3.117$ is detected in X-ray 
photometry at 
the narrow-band source position.
At $z=3.1$, Chandra receives X-ray emission from 2-30 keV (rest-frame), 
meaning that Compton thick obscuration ($N_\mathrm{H}>\sim 10^{23}$cm$^{-2}$)
is needed to hide AGN.  Even such heavily obscured AGN are likely to 
reveal their presence via narrow emission lines, which should 
be indicated by high-ionization UV 
lines like C IV accompanying Ly $\alpha$.  
Amongst our LAE spectra, only the three X-ray sources 
show signs of AGN activity in the form of emission lines other than 
Ly $\alpha$, and the other
159 LAEs are undetected in a stacked X-ray image 
\citep{gronwalletal07}.
We therefore expect that very few LAEs contain 
AGN that dominate their Ly$\alpha$ or continuum emission.  
Our two X-ray sources at $z=3.1$ imply that AGN are present in 
only $1.2\pm0.9$\% of Ly $\alpha$ selected galaxies at this redshift.
We restrict our subsequent analysis to the 159 LAEs without X-ray detections.

\section{CLUSTERING ANALYSIS}
\label{sec:clust}

We used the \citet{landys93} estimator to measure the angular 
correlation function using histograms of pairs of points at 
separation $\theta$ between the data catalog ($D$) and 
itself ($DD$) as well as the cross-correlation and autocorrelation 
with a set of random catalogs ($R$).  
We modelled this observed correlation function as an intrinsic 
$w(\theta)$ minus the ``integral constraint'' caused by estimating the 
sky density of LAEs from our own survey 
\citep[see][H.~Francke et al. in prep]{peebles80,infante94}

\begin{equation}
\frac{DD(\theta) - 2DR(\theta) + RR(\theta)}{RR(\theta)} =
w(\theta) -\sigma^2 (1+w(\theta))
 \; \; \; ,
\label{eq:ls-modified}
\end{equation}
\begin{equation}
\mathrm{where}~~~ \sigma^2 = \frac{1}{\Omega^2} \int\!\!\!\int_{\Omega} 
w(\theta_{12}) d \Omega_1 d\Omega_2 
 \; \; \; .
\label{eq:ic}
\end{equation}

\begin{figure}[h!]
\epsscale{1.2}
\plotone{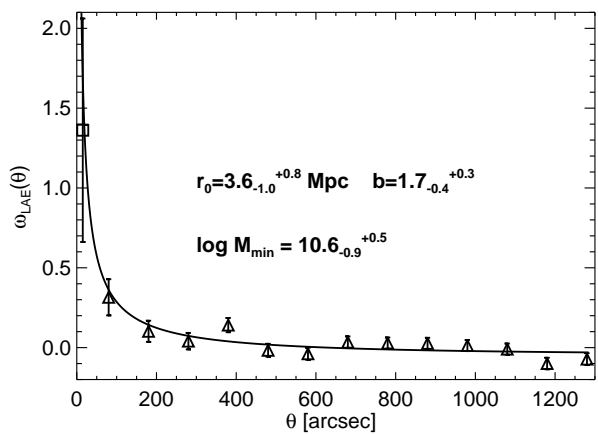}
\caption[]
{ 
Angular auto-correlation function data with best-fit 
model (solid line) and best-fit 
spatial correlation length and dark matter halo bias and 
mass parameters listed.  The bin below 30'' (square) is not used in the 
fit.  
}
\label{fig:wtheta}
\end{figure}

\begin{figure}[h!]
\epsscale{1.2}
\plotone{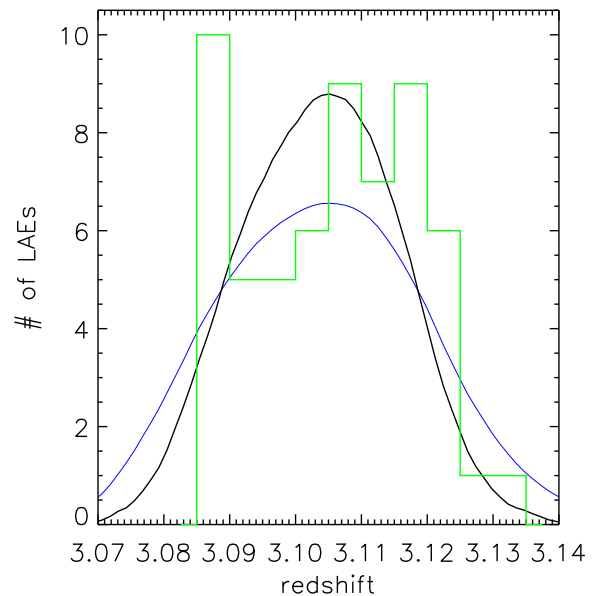}
\caption[]
{
Histogram of LAE redshifts in ECDF-S.  The thin solid curve is the 
filter transmission curve normalized to have the same area, and the 
thick solid curve is the expected redshift distribution described 
in the text.
}
\label{fig:zhist}
\end{figure}

Figure \ref{fig:wtheta} 
shows our binned data along 
with the best-fit model.  
Error bars show the uncertainties in each 
bin of $\sigma_w(\theta) = (1+w(\theta))/\sqrt{RR(\theta)}$ 
\citep{landys93,gawiseretal06a}.  
The analysis was restricted to scales larger than 30$''$ 
($\approx$1 comoving Mpc) 
that are insensitive to the possible presence of 
multiple LAEs in some dark matter halos.
In order to determine the  
expected redshift distribution, $N_{exp}(z)$, 
we performed a Monte Carlo simulation placing a large number of LAEs 
over the redshift range $3<z<3.2$ and assigning them equivalent widths 
drawn at random from the equivalent width distribution observed 
for our sample \citep{gronwalletal07}.  We then used the   
NB4990 filter bandpass to calculate the excess narrow-band flux that 
would be observed, removed all LAEs with 
``observed'' equivalent width <80{\AA}, and measured the redshift 
distribution of the selected objects.  
Fig. \ref{fig:zhist} shows that the  
expected redshift distribution is narrower than the transmission 
curve.  This occurs because 
LAEs far from the central redshift must have very high 
equivalent widths to be selected through narrowband excess, 
due to the reduced filter transmission
at the wavelength of their Ly $\alpha$ emission lines.  
Fig. \ref{fig:zhist} shows that the 
histogram of observed LAE redshifts is consistent with $N_{exp}(z)$.
The only bin inconsistent with poisson fluctuations of the expected 
redshift distribution at the 9%
is at $3.085 \leq z < 3.090$, which displays a 
$3\sigma$ excess, revealing an overdensity at this redshift.

We used $N_{exp}(z)$ to 
deproject the angular correlation function,  
following \citet{simon07}.
In order to avoid degeneracy between the clustering length $r_0$ and the 
power-law index $\gamma$ in the underlying  
spatial correlation function, 
$\xi(r)=(r/r_0)^{-\gamma}$,  
we assumed a typical power-law with $\gamma=1.8$.
This yielded a moderate clustering length, 
$r_0=3.6^{+0.8}_{-1.0}$ Mpc (comoving).  
The narrow redshift distribution of narrow-band 
selected LAEs reduces the loss of angular clustering signal due to 
projection, allowing a  
high S/N measurement of moderate clustering.  
Following \citet{quadrietal07a}, 
this value of $r_0$ corresponds to 
stronger clustering than the dark matter at $z=3.1$ by 
a bias factor $b=1.7^{+0.3}_{-0.4}$.
Note that bias factors are robust to the degeneracy between 
$r_0$ and $\gamma$.  
This bias factor is shared by the population of dark matter halos 
with masses greater than  
$\log_{10}\mathrm{M_{min}} 10.6^{+0.5}_{-0.9}$ M$_\odot$
\citep{shetht99}, implying a median dark matter halo mass of 
$\log_{10}\mathrm{M_{med}} = 10.9^{+0.5}_{-0.9}$ M$_\odot$.
If 10\% of the LAEs were unclustered low-redshift contaminants, the corrected 
value of $r_0$ would be 10\% higher, yielding halo masses $\sim50$\% higher.  

The comoving number density of our LAE sample is 
$1.5\pm0.3\times10^{-3}$Mpc$^{-3}$ \citep{gronwalletal07}, 
where the uncertainty includes variance 
due to large-scale structure in our survey volume 
for objects with $b=1.7$ \citep{somervilleetal04}.  
The number density of the corresponding dark matter halos is 
$3^{+25}_{-2}\times10^{-2}$Mpc$^{-3}$, 
implying a
``mean halo occupation'' of $5^{+10}_{-4.5}$\% for the LAEs.
There is significant freedom in how the LAEs 
could be assigned to this subset of the available dark matter halos.
However, the LAE median halo mass must roughly follow the result 
$\log_{10}\mathrm{M_{ med}} = 10.9^{+0.5}_{-0.9}$ M$_\odot$ 
in order to explain the observed clustering 
bias.\footnote{The quantity that must 
be preserved is the mean halo bias.  The median halo mass is 
a simpler statistic that is also robust in typical HOD models, and the 
difference is far smaller than the reported uncertainties.}  Explorations of 
complex halo occupation distribution (HOD) models show that the assumption 
of one galaxy per halo made in our determination of M$_\mathrm{med}$ 
can cause additional uncertainties of up to 0.2dex at $z=3.1$,$b=1.7$ 
\citep{leeetal06,zhengetal07}.

\section{SED MODELLING}
\label{sec:sed}

\begin{figure}[h!]
\epsscale{1.2}
\plotone{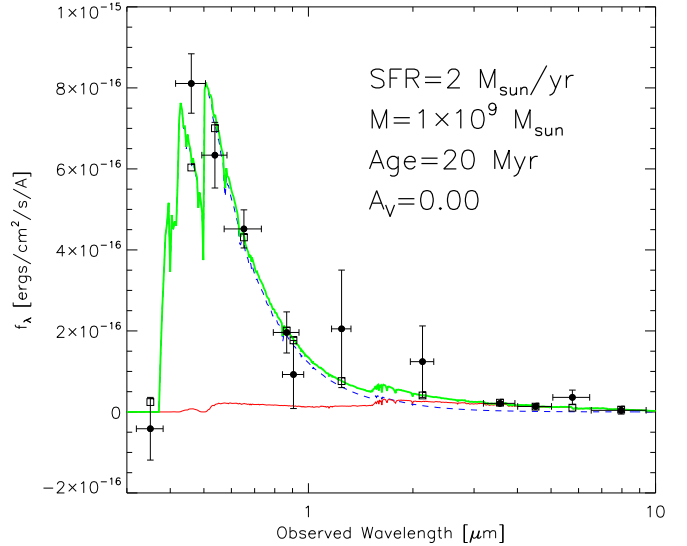}
\caption[]
{
Datapoints show  stacked flux densities ($f_\lambda$) of LAEs lacking individual detections in 
the SIMPLE IRAC images,  
with 1$\sigma$ error bars.  
The thick solid curve 
gives the best-fit model described in the text, which is a sum of 
a young component (dashed curve) and an old component (thin solid 
curve).  Squares show fluxes predicted by the best-fit model, which 
 has $\chi^2/\mathrm{d.o.f.}=14.6/12$. 
}
\label{fig:sed}
\end{figure}

\citet{laietal07b}
offer a detailed description of our IRAC 
photometry along with single-component SED fitting of the detected and 
undetected objects and a comparison of their continuum properties 
with those of Lyman break galaxies.
76 of our LAEs fall within regions of the 
SIMPLE images (which include the GOODS IRAC images)
where the lack of bright neighbors enables 
IRAC photometry accurate to very low fluxes.  
Only 
24 LAEs
(30\%)
are detected by IRAC at fluxes above the 
$2 \sigma$ SIMPLE flux limit;  these objects represent the high-mass end 
of the LAE mass function and appear to have stellar masses $>3\times10^9$M$_\odot$  
Only 2 of these 
LAEs are detected in our $J,K$ images, which are two magnitudes 
shallower than the IRAC 3.6,4.5 $\mu$m images. 
The IRAC-detected LAEs are brighter in the rest-UV and rest-optical 
continuua, with mean R-band and 3.6$\mu$m fluxes 
corresponding to magnitude 25.4 and 24.4 respectively,  
compared with 26.7 and 26.6 for LAEs not detected by IRAC.  
In order to investigate the full SED of typical LAEs, which are too dim 
to be detected individually in our NIR and Spitzer images, we 
measured average fluxes from stacked images of the 
52 LAEs
(70\%)
lacking IRAC detections.  
We show the resulting SED in Fig. \ref{fig:sed}, where the 
$V$-band flux has been corrected for the contribution of the Ly$\alpha$ 
emission lines to this filter.     
Uncertainties in the stacked photometry were determined using bootstrap 
resampling to account for both sample variance and photometric errors.

\begin{figure}[h!]
\rotatebox{90}{
 \scalebox{0.75}[0.75]{
\includegraphics{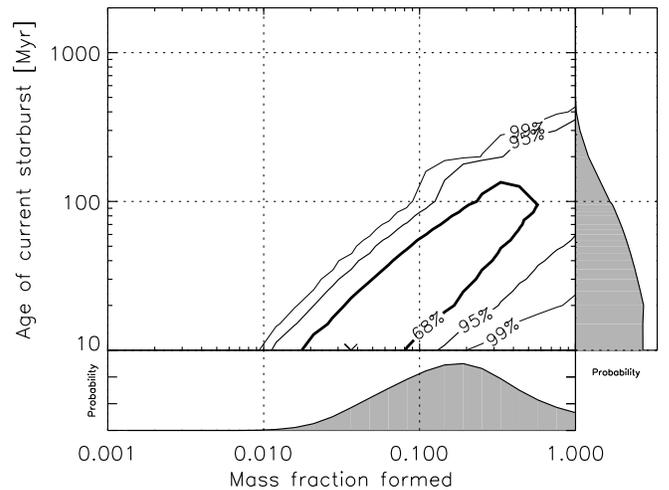}
}
}
\caption[]
{
Constraints on age of the young stellar population versus its mass fraction.
}
\label{fig:agemass}
\end{figure}

Instead of modelling LAEs with a single stellar population,  
we analyzed the extent to which the data allow the presence 
of an underlying evolved population.  
We adapted the method of
K.~Schawinski et al. (in prep.)
to model the star formation histories
using a two-burst scenario with the old component as an instantaneous
burst and the young component as an exponentially declining starburst
with variable e-folding time. 
\citet{maraston05} 
population synthesis models were used with 
metallicity ranging from 0.02 solar to solar, a 
\citet{salpeter55}
initial mass function,
and the \citet{calzettietal00} dust law.  
The best-fit model shown in  Fig. \ref{fig:sed} 
corresponds to a stellar mass of 
$1.0^{+0.6}_{-0.4}\times10^9$ M$_\odot$, star formation rate 
of $2\pm1$M$_\odot$yr$^{-1}$, 
 and dust extinction $A_V=0.0^{+0.1}_{-0.0}$ (only positive values 
of $A_V$ were considered).   
Figure \ref{fig:agemass} shows the results for the age of the young 
population versus the mass fraction of the young population.
The young population has an age of $20^{+30}_{-10}$Myr with an 
e-folding timescale $\tau=750\pm250$Myr i.e., a nearly 
constant star formation rate.    
Although we did not include our narrow-band photometry in the SED 
analysis, the median LAE rest-frame equivalent width of 60{\AA} 
found by \citet{gronwalletal07} is consistent with that expected for 
normal stellar populations in this age range 
\citep{finkelsteinetal07}. 
The age of the old population is not well constrained but has a 
best fit of 2 Gyr (the age of the universe at $z=3.1$).

The mass fraction formed in the current starburst is not well constrained, 
and models where all of the stellar mass was
produced in a current burst of star formation of age 60 to 350 Myr are allowed.
Indeed, \citet{laietal07b}
performed a one-component SED fit with \citet{maraston05} 
models and found a best-fit age of 100 Myr, $\tau$=250 Myr, E(B-V)=0, and 
M$_*=3\times10^8$M$_\odot$.  Our two-component best-fit is preferred to 
this, even accounting for the two extra degrees of freedom, but a single 
``$\tau$-model'' population is not ruled out at 95\% confidence (see Fig. 5).  

\section{DISCUSSION}
\label{sec:discussion}

In CDM cosmology, galaxy formation is an ongoing process caused 
by merging of lower-mass dark matter halos, which may already possess 
stars.
Finding stellar population ages of $<100$ Myr is interesting.
Our analysis of halo merger trees from the Milli-Millenium 
simulation \citep{springeletal05} found the 
median age of dark matter halos with M$>10^{10.6}$M$_\odot$ at $z=3.1$ 
(defined as the age since half of the dark matter mass was accumulated)
to be $\sim$600 Myr, 
with only $<$10\% of halos younger than 100 Myr. 
If repeated LAE phases occur, 
the mean halo occupation of $\sim1$--$10$\% can be interpreted as a 
``duty cycle'' telling us what fraction of each halo's lifetime is 
spent in the early phases of starbursts before significant dust 
is generated, and the population-averaged young age of $\sim$20 Myr would 
imply that this phase typically lasts $\sim$40 Myr.      
Alternatively, 
if all dark matter halos experience a single LAE phase shortly after their 
``formation'' in a major merger, the mean halo occupation implies 
that LAEs will only be found in the youngest $1$--$10$\% 
of halos, which is barely consistent with their single-population best-fit 
age of 100 Myr.  
If LAEs represent a subset of dark matter halos selected to have 
ages less than 100 Myr, their observed clustering may 
underestimate their dark matter halo masses by up to a factor of two 
\citep[see][]{gaow07}.   

\begin{figure}[h!]
\epsscale{1.25}
\plotone{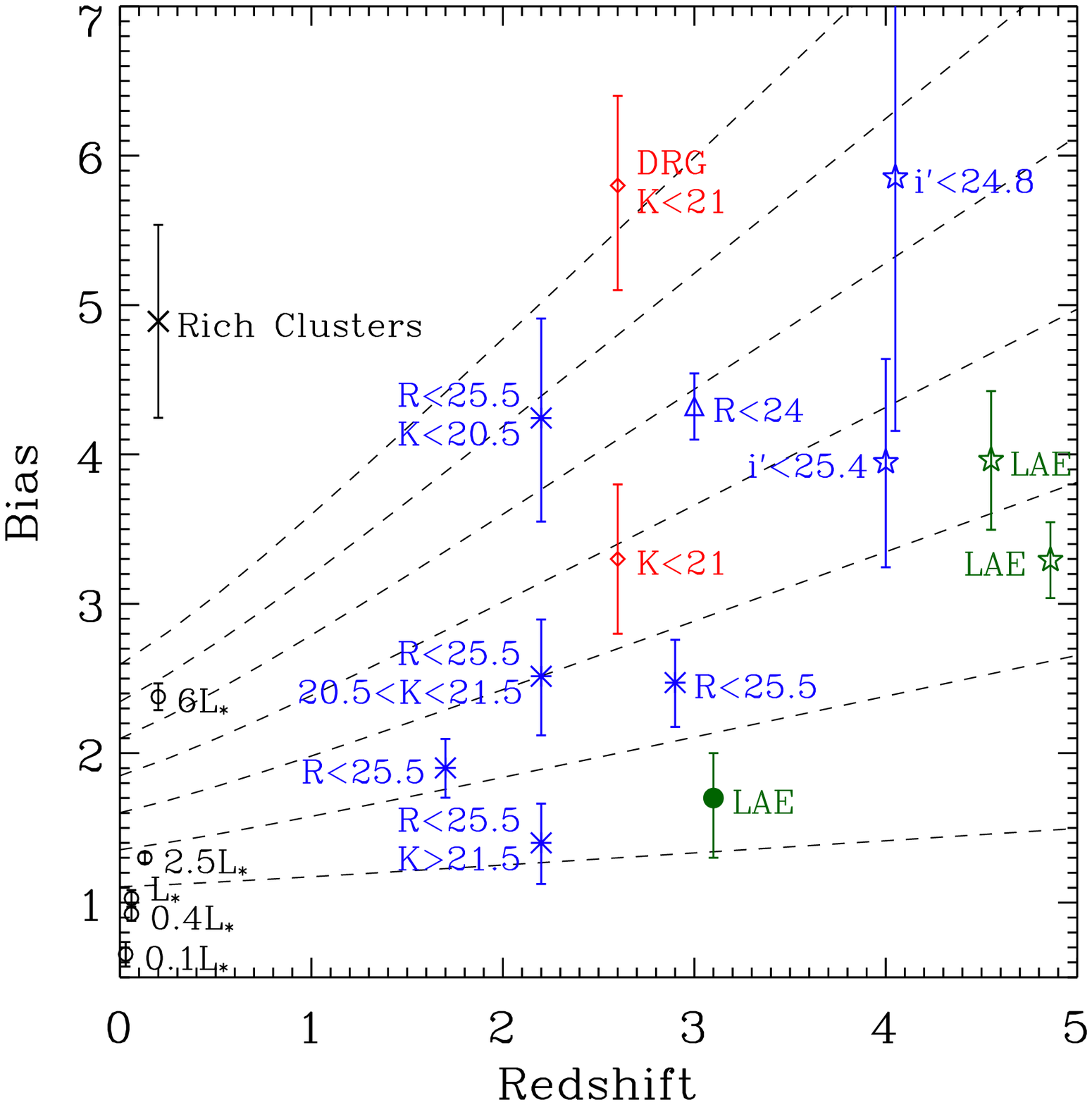}
\caption[]
{ 
Tracks show the evolution of bias with redshift calculated 
using the no-merging model.  The filled circle shows
our result for the bias of LAEs at $z=3.1$.  
Previous results at high-redshift are shown for 
LAEs at $z=4.5$ and $z=4.86$ (stars, \citealp{kovacetal07} and 
\citealp{ouchietal03}, respectively), 
LBGs at $z\sim4$ (stars, \citealp{ouchietal04b}),  
K-selected galaxies (diamonds, \citealp{quadrietal07a}),   
bright LBGs at $z\sim 3$ (triangle, \citealp{leeetal06}),   
and BM, BX, and LBG galaxies (asterisks, \citealp{adelbergeretal05b},
\citealp{adelbergeretal05a}).  
Local galaxy clustering is shown for SDSS galaxies 
(open circles, \citealp{zehavietal05})
and for rich clusters (cross, \citealp{bahcalletal03}). 
K-band limits are in Vega magnitudes. 
}
\label{fig:bias}
\end{figure}

Figure \ref{fig:bias} shows the reported bias values for LAEs to be lower 
than those of other $z>3$ galaxy populations
\citep[bias values determined as in][]{quadrietal07a}.  
  The expected evolution of 
bias is shown for the ``no-merging'' model \citep{fry96,whiteetal07}.  
A realistic amount of 
merging will cause the bias to drop somewhat faster, so the plotted 
trajectories provide an upper limit 
on the bias factor of a given point at lower redshifts.  This shows 
that typical $z=3.1$ LAEs 
will evolve into galaxies of at most a few times $L^*$ 
at $z=0$.  
The bias values imply  
that LAEs at $z=3.1$ might evolve into 
the subset of BX galaxies at $z\simeq 2.2$ dimmer than $K=21.5$, which 
also show relatively weak clustering \citep{adelbergeretal05a}.  
The $K>21.5$ BX galaxies have average M$_*=1.5\times10^{10}$M$_\odot$, 
so the $z=3.1$ LAEs would need to form stars at an average rate of 
14 M$_\odot$yr$^{-1}$ over the intervening Gyr.  This could be achieved 
with a constant specific SFR and no merging or 
with a constant SFR and $\sim$2 major mergers.  The only previous 
measurements of LAE clustering 
in unbiased fields are at $z=4.5$ \citep{kovacetal07} and 
$z=4.86$ \citep{ouchietal03}, 
and this earlier LAE population appears to have significantly stronger 
clustering, consistent with possibly evolving into typical Lyman break 
galaxies at $z\simeq 3$.  

The models of \citet{ledelliouetal06} predict stellar and dark matter 
masses and star formation rates for $z=3$ LAEs within a factor of two of 
our results, despite assuming a top-heavy IMF and a very low 
escape fraction $f_{esc}=0.02$ that appears 
inconsistent with the observed lack of dust
\citep[see][for an alternative approach]{kobayashietal07}.  
\citet{maoetal07} used the stellar mass of $5\times10^8$M$_\odot$ 
observed by \citet{gawiseretal06b} to predict LAE dark matter masses of 
$10^{10}<$M$<10^{11}$M$_\odot$, in the lower end of our allowed range. 
 The stellar ages of $\sim 20$ Myr preferred by the two-population fit are  
noticeably 
lower than the maximum values of 100 to 500 Myr predicted by these authors, 
\citet{moriu06}, and \citet{haimans99}, but the 
ages of 60 to 350 Myr preferred for the case of no evolved stars would 
be compatible.

\begin{figure}[h!]
\epsscale{1.25}
\plotone{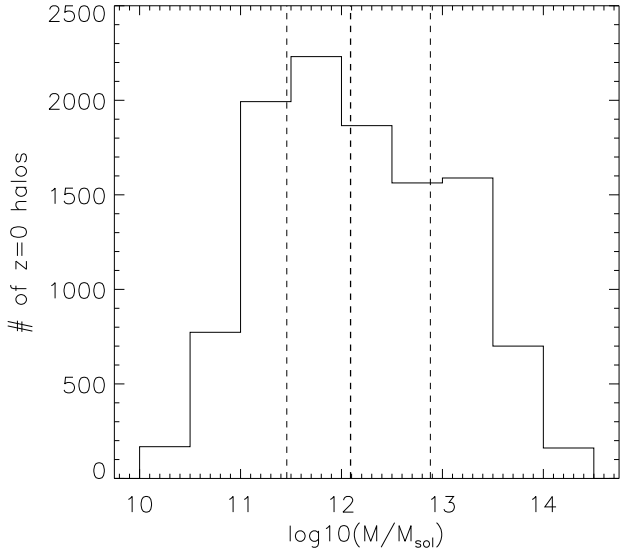}
\caption[]
{ 
Histogram of dark matter halo masses of present-day descendants of halos 
with M$>5\times10^{10}$M$_\odot$ at $z=3.1$.  The dashed lines show the 
median halo mass of $1.2\times10^{12}$M$_\odot$ and the 25th and 75th 
percentile values of  $2.9\times10^{11}$M$_\odot$  and 
$7.6\times10^{12}$M$_\odot$.
}
\label{fig:mass}
\end{figure}

None of the current models and numerical simulations of LAEs 
\citep[see also][]{thommesm05,razoumovs06,tasitsiomi06}
predict their present-day descendants.
Nonetheless, the evolution of a significant fraction of $z=3.1$ 
LAEs into $z=0$ $L^*$ galaxies with 
dark matter mass  $M_{DM}\simeq2\times10^{12}$M$_\odot$ 
and stellar mass $M_*\simeq4\times10^{10}$M$_\odot$ 
\citep{ichikawaetal07} 
appears reasonable. 
Fig. \ref{fig:mass} shows the histogram of present-day masses of 
dark matter halos in the Milli-Millenium merger trees that have 
progenitors with M$>5\times10^{10}$M$_\odot$ at $z=3.1$.  The median 
present-day halo mass is $1.2\times10^{12}$M$_\odot$, and this would increase 
if LAEs found in sub-halos of massive $z=3.1$ halos were included.  
\citet{lietal07} predict that the main progenitor of a present-day 
$L^*$ galaxy had a dark matter mass of 
$\sim10^{11}$M$_\odot$ at $z=3$ and that these 
galaxies experienced several major mergers at $1.5<z<7$.  
To form an $L^*$ galaxy at $z=0$, 
several LAEs could merge while experiencing a mild reduction in average SFR, 
with accretion of lower-mass dark matter halos through minor mergers 
providing most of the final dark matter mass.  
However, the halo mass distribution of $z=0$ descendants in Fig.\ref{fig:mass}
is very broad, with 25th and 75th percentile values of $2.9\times10^{11}$M$_\odot$ and $7.6\times10^{12}$M$_\odot$.  While $z=0$ $L^*$ galaxies like the 
Milky Way are roughly the median descendants of $z=3.1$ LAEs, the descendant
halos include a wide range from dwarf galaxies to rich galaxy groups.

The typical LAE stellar mass 
at $z=3.1$ 
is lower than that of any other 
studied high-redshift population \citep[see][]{reddyetal06b}
but is close to that of dim ($i<26.3$) Lyman break galaxies (LBGs) 
at $z\sim5$ \citep{vermaetal07}. 
LAEs at $z=3.1$ 
have much lower star formation rate, stellar age, 
stellar mass, dark matter halo mass, and dust extinction than 
the $\sim30$M$_\odot$yr$^{-1}$, $\sim500$ Myr, $\sim2\times10^{10}$M$_\odot$,
 $\sim3\times10^{11}$M$_\odot$,
 $A_V\simeq1$ LBG
population at $z\sim3$ 
\citep[$R<25.5$,][]{shapleyetal01,adelbergeretal05a} or the 
 $\sim100$M$_\odot$yr$^{-1}$, $\sim2$ Gyr, $\sim10^{11}$M$_\odot$, 
 $\sim10^{13}$M$_\odot$,
$A_V\simeq2.5$ Distant Red Galaxy (DRG)
population \citep{webbetal06,forsterschreiberetal04,quadrietal07a}.  
The high-redshift  Sub-Millimeter Galaxies
\citep{chapmanetal03b}
appear to be 
the most massive and dusty, with the highest SFR.
LAEs may 
represent the beginning of an evolutionary sequence where galaxies
gradually become more massive and dusty due to mergers and
star formation, but most LAEs at $z=3.1$ will never reach the 
DRG stage since DRG stellar and dark matter 
masses are already greater than those 
of present-day $L^*$ galaxies.  

The Damped Ly$\alpha$ Absorption systems (DLAs, \citealp{wolfeetal05}) 
are another high-redshift population that probes the faint end of 
the luminosity function.
The dark matter halo masses of DLAs at $z\sim3$  were determined by 
\citet{cookeetal06b} to lie in the range $10^9<$M$<10^{12}$M$_\odot$ 
i.e.,  
$1.3<b<4$, 
which overlaps with the range of both $L^*$ and super-$L^*$ progenitors 
in Fig. \ref{fig:bias}.  
At least half of 
the DLAs appear to have ongoing star formation \citep{wolfeetal04} 
and two of the three DLAs detected in emission were 
seen in Ly$\alpha$.  Further study 
is needed to determine 
the relationship between DLAs and LAEs.

The observed properties of LAEs at $z=3.1$ make them the most promising 
candidates to be high-redshift progenitors of present-day $L^*$ galaxies like
the Milky Way.  
Our results suggest that LAEs are observed during the early phases of a 
burst of star formation, perhaps caused by a major merger of smaller 
dark matter halos.  The input halos appear to have already contained 
stars, accounting for the evolved stellar population that appears to 
contribute most of the LAE stellar mass, although starburst-only models 
are also allowed.  
 It is clear that not all progenitors of L$^*$ galaxies 
were LAEs at $z=3.1$.
The comoving number density of our sample of LAEs 
is a factor of 15 less than $\phi^*$ for local galaxies \citep{linetal96}, 
plus we expect several high-redshift halos to merge into a single 
galaxy today.  
It remains possible that all progenitors of present-day galaxies 
experienced an LAE phase at {\it some} redshift.  
Clustering and SED studies 
of LAEs at various redshifts are needed to 
assess the validity of this hypothesis.

\acknowledgments

We acknowledge valuable conversations with  
Kyoung-Soo Lee, Jeff Newman, Ravi Sheth, David Spergel, Jason Tumlinson 
and Martin White.  
We are grateful for 
support from Fundaci\'{o}n Andes, the FONDAP Centro de Astrof\'{\i}sica, 
and the Yale Astronomy Department.
Support for this work was provided by NASA through an award issued by 
JPL/Caltech.
This material is based upon work supported by the National Science 
Foundation under Grant. Nos. AST-0201667, 
an NSF Astronomy and Astrophysics Postdoctoral 
Fellowship (AAPF) awarded to E.G., and AST-0137927 awarded to R.C.   
We thank the staff 
of Cerro Tololo Inter-American Observatory and 
Las Campanas Observatory for their invaluable 
assistance with our observations.  
The Millenium Simulation databases used in this paper and the web application 
providing online access to them were constructed as part of the 
activities of the German Astrophysical Virtual Observatory.  
This research has made use of NASA's Astrophysics Data System.  
Facilities:CTIO(MOSAIC II),LCO(IMACS)

\end{document}